\documentclass{WileyMSP-template}

\usepackage{bm}
\usepackage[usenames,dvipsnames]{xcolor}
\usepackage{srcltx}
\usepackage{graphicx}
\usepackage{epstopdf}
\usepackage{color}
\usepackage{amsmath}
\usepackage{amssymb}
\usepackage{color}
\usepackage{float}
\usepackage{subfigure}
\usepackage{gensymb}

\usepackage{epstopdf}
\usepackage{epsfig}
\usepackage{psfrag}

\usepackage{cuted}
\usepackage[mode=errorstop]{pstool}
    \newcommand{\idt}{\hspace{5mm}}
    \newcommand{\figref}{Figure~\ref}

\begin{document}

\pagestyle{fancy}

\title{Generalized Brewster Effect \\ using Bianisotropic Metasurfaces}

\maketitle

% Author: Please give full first and last names for authors and include * after the name of all corresponding authors
\author{Guillaume Lavigne*, Christophe Caloz}

% Affiliations: Please provide adacemic titles (Prof. or Dr.) for all authors where applicable, and include an institutional email address for all corresponding authors
\begin{affiliations}
G. Lavigne\\
Electrical Engineering
Polytechnique Montr\'eal \\
Montr\'eal H3T 1J4, CA \\
Email: guillaume.lavigne@polymtl.ca\\
Prof. C. Caloz\\
Faculty of Engineering Science \\
KU Leuven \\
Leuven 3000, BE \\
\end{affiliations}

% Keywords: Please provide a minimum of three and a maximum of seven keywords, separated by commas


% Abstract should be written in the present tense and impersonal style (i.e., avoid we), and be at most 200 words long
\begin{abstract}
We show that a properly designed bianisotropic metasurface placed at the interface between two arbitrary different media, or coating a dielectric medium exposed to the air, provides Brewster (reflectionless) transmission at arbitrary angles and for both the TM and TE polarizations. We present a rigorous derivation of the corresponding surface susceptibility tensors based on the Generalized Sheet Transition Conditions (GSTCs), and demonstrate the system with planar microwave metasurfaces designed for polarization-independent and azimuth-independent operations. Moreover, we reveal that such a system leads to the concept of effective refractive media with engineerable impedance. The proposed bianisotropic metasurfaces provide deeply subwavelength matching solutions for initially mismatched media, and alternatively lead to the possibility of on-demand manipulation of the conventional Fresnel coefficients. The reported generalized Brewster effect represents a fundamental advance in optical technology, where it may both improve the performance of conventional components and enable the development of novel devices.
\end{abstract}

% Text: Please use section headings and subheadings as specified below. For communications, all section headings apart from Experimental Section should be removed
% Please make the first reference to a display item bold: \textbf{Figure 1}
% Do not abbreviate Figure, Equation, etc.; display items are always singular, i.e., Figure 1 and 2.
% Equations are always singular, i.e., Equation 1 and 2, and should be inserted using the {equation} environment, not as graphics
% Please do not use footnotes in the text, additional information can be added to the Reference list.

\vspace{1cm}

\idt The Brewster effect, which consists in the vanishment of the reflection of TM-polarized waves at the interface between two dielectric media at a specific incidence angle~\cite{saleh1991fundamentals}, has a history of more than 200 years. In 1808, Malus observed that unpolarized light becomes polarized upon reflection under a particular angle a the surface of water~\cite{malus1809propriete}. Seven years later, Brewster experimentally showed that this angle was equal to the inverse tangent of the ratio the refractives indices of the two media~\cite{brewster1815laws}. Another six years later, in 1821, Fresnel completed the understanding of the phenomenon using a mechanical model of the interface system and derived the eponymic reflection and transmission coefficients~\cite{fresnel1834memoire}, which embed the Brewster effect. Finally, these formulas were generalized to magneto-electric materials, which support either TM-polarization or TE-polarization Brewster transmission, with both possible only for normal incidence, by Giles and Wild~\cite{giles1985brewster}.

\idt The recent advent of metasurfaces has created novel opportunities to extend the Brewster effect. Metasurfaces allow indeed unprecedented manipulations of electromagnetic waves~\cite{glybovski2016metasurfaces,achouri2018design}; specifically, bianisotropic metasurfaces~\cite{asadchy2018bianisotropic} may produce full polarization transformation~\cite{pfeiffer2014bianisotropic}, anomalous reflection~\cite{asadchy2015functional} and diffractionless generalized refraction~\cite{lavigne2018susceptibility,wong2018perfect}. They have recently been shown to support Brewster-like, i.e., reflection-less, transmission when surrounded at both sides by air in planar optical silicon nanodisk  configuration~\cite{paniagua2016generalized} and in non-planar microwave split-ring resonator configuration~\cite{tamayama2015brewster,yin2019metagrating}. Moreover, they have been demonstrated to allow general Brewster transmission, i.e., between two different media, in the particular case of normal incidence in a non-planar bianisotropic loop-dipole configuration~\cite{dorrah2018bianisotropic}.

\idt Here, following our initial suggestion in~\cite{lavigne2018extending}, we present a generalization of the Brewster effect between two arbitrary different media, for arbitrary incidence angle and arbitrary polarization, using a planar bianisotropic metasurface. We derive synthesis formulas of the corresponding metasurface susceptibility tensors and demonstrate the generalized Brewster angle by full-wave electromagnetic simulation.

\begin{figure}[h]
\centering
\includegraphics[width=\linewidth]{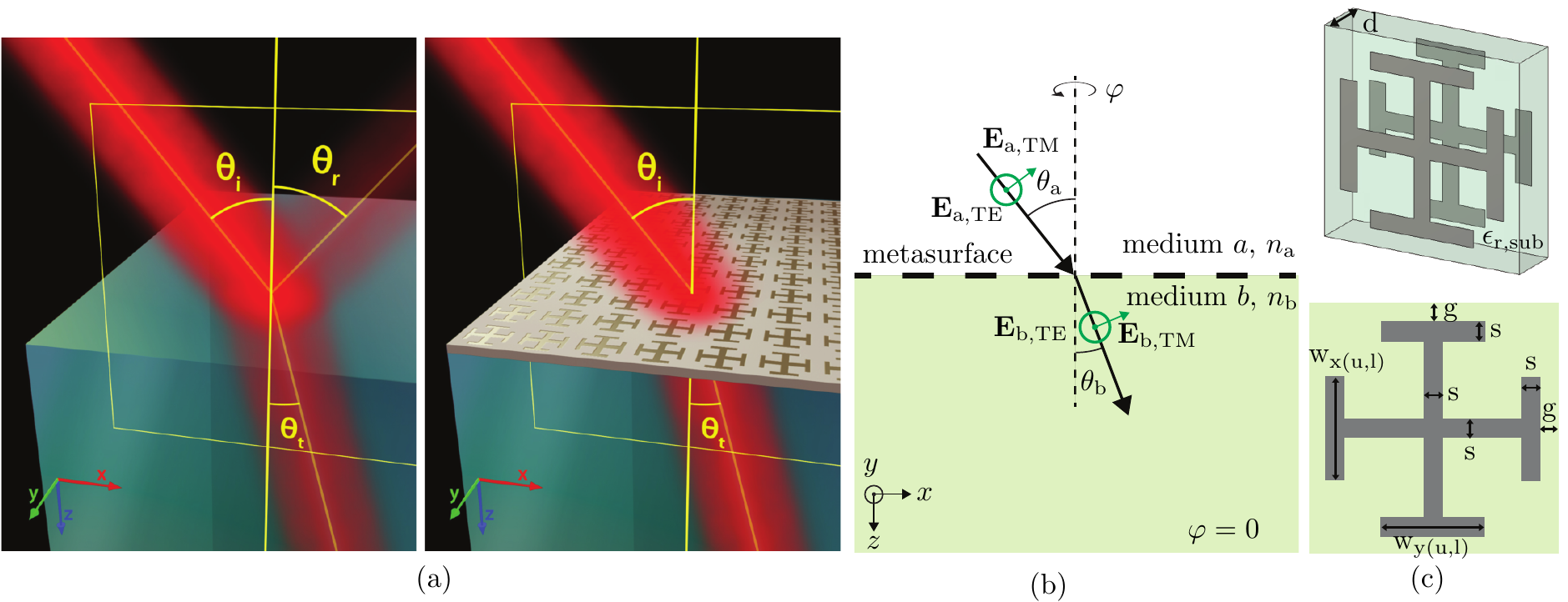}
\caption{Metasurface-based generalized Brewster refraction between two media. (a)~Scattering of a wave impinging on the interface under an arbitrary angle ($\theta_\text{a}$), with conventional Fresnel transmission and reflection for the case of the bare interface (left) and with reflectionless (Brewster) transmission when a properly designed metasurface is placed at the interface (right). (b)~Brewster metasurface problem, for TE and TM polarizations, in the $xz$-plane scattering plane. (c)~Proposed 2-layer conducting cross-potent metaparticle for a microwave proof of concept.}\label{fig:Brewster_problem}
\end{figure}

\idt Figure~\ref{fig:Brewster_problem} shows the proposed metasurface-based generalized Brewster effect. Figure~\ref{fig:Brewster_problem}(a) illustrates the suppression of reflection for arbitrary wave incidence angle and arbitrary polarization, \figref{fig:Brewster_problem}(b) defines the corresponding problem in the plane of scattering, and \figref{fig:Brewster_problem}(c) depicts the metaparticule used in the paper as a proof of concept in the microwave regime. The metasurface is assumed to suppress reflection without altering the direction of refraction prescribed by the Snell law for the initial pair of media and without inducing any gyrotropy (polarization rotation), while being passive, lossless and reciprocal. The preservation of the Snell law for the transmitted wave implies a uniform (without phase gradient) metasurface, and hence a uniformly periodic metastructure.

\idt We consider the metasurface problem depicted in~\figref{fig:Brewster_problem}~(b), where a wave incident from the medium $a$ in the $xz-$plane at an arbitrary angle $\theta_\text{a}$ is fully transmitted, without reflection ($R=0$), into the medium $b$, at the Snell angle $\theta_\text{b} = \arcsin (\frac{n_\text{a}}{n_\text{b}}\sin \theta_\text{a}) $. We assume the time-harmonic complex convention $\exp(-i\omega{t})$ throught the paper, and we shall apply the metasurface synthesis technique described in~\cite{achouri2020electromagnetic} to determine the susceptibility tensors of the metasurface.

The first step in the synthesis is to define the desired tangential fields at both sides of the metasurface in the plane $z=0$. For a TM-polarized wave, these fields read

\begin{subequations}\label{eq:fields_TM}
\begin{equation}
\mathbf{E}_{\parallel a,\text{TM}}=\cos\theta_\text{a}e^{-ik_{\text{a}x}x}\hat{x},\quad
\mathbf{H}_{\parallel a,\text{TM}}=\frac{e^{-ik_{\text{a}x}x}}{\eta_\text{a}}\hat{y},
\end{equation}

\begin{equation}
\mathbf{E}_{\parallel b,\text{TM}}=T\cos\theta_\text{b}e^{-ik_{\text{b}x}x}e^{i \phi_\text{TM}}\hat{x},\quad
\mathbf{H}_{\parallel b,\text{TM}}=T\frac{e^{-ik_{\text{b}x}x}}{\eta_\text{b}}e^{i \phi_\text{TM}}\hat{y},
\end{equation}
\end{subequations}

while for a TE-polarized wave, they read

\begin{subequations}\label{eq:fields_TE}
\begin{equation}
\mathbf{E}_{\parallel a,\text{TE}}=-e^{-ik_{\text{a}x}x}\hat{y},\quad
\mathbf{H}_{\parallel a,\text{TE}}=\cos\theta_\text{a}\frac{e^{-ik_{\text{a}x}x}}{\eta_\text{a}}\hat{x},
\end{equation}
\begin{equation}
\mathbf{E}_{\parallel b,\text{TE}}=-Te^{-ik_{\text{b}x}x}e^{i \phi_\text{TE}}\hat{y},\quad
\mathbf{H}_{\parallel b,\text{TE}}=T\cos\theta_\text{b}\frac{e^{-ik_{\text{b}x}x}}{\eta_\text{b}}e^{i \phi_\text{TE}}\hat{x},
\end{equation}
\end{subequations}
where $k_{(a,b)x}$ is the tangential wavenumber, $\eta_{(a,b)}$ is the wave impedance in medium $(a,b)$, and $T$ is the transmission coefficient between the two media. The phase terms $e^{i \phi_\text{TM,TE}}$ in the transmitted fields are introduced here to account for the typical phase shifts imparted to the wave by a pratical metasurface and to provide degrees of freedom that may be advantageous in the design of the unit-cell metaparticle. In these relations, $T$ is obtained by enforcing power conservation across the metasurface (passivity and losslessness assumptions), i.e., by enforcing the continuity of the normal component of the Poynting vector at $z=0$~\cite{achouri2020electromagnetic}. This leads, using the fields in~\eqref{eq:fields_TM} and~\eqref{eq:fields_TE}, to
\begin{equation}\label{eq:transmission_coefficient_T}
  T =\sqrt{\frac{\eta_\text{b}\cos\theta_\text{a}}{\eta_\text{a}\cos\theta_\text{b}}},
\end{equation}
which is identical for the TE and TM polarizations.

\idt The boundary conditions in the plane of the metasurface ($z=0$) are the general sheet transition conditions~(GSTCs)~\cite{achouri2020electromagnetic}
\begin{subequations}\label{eq:GSTC_tangential}
	\begin{equation}
		\hat{z} \times \Delta\mathbf{H} =  i \omega \epsilon \overline{\overline{ \chi}}_\text{ee} \mathbf{E}_\text{av} +  i \omega \overline{\overline{ \chi}}_\text{em} \sqrt{\mu \epsilon}  \mathbf{H}_\text{av} ,
	\end{equation}
	\begin{equation}
		\Delta \mathbf{E} \times \hat{z}   = i \omega \mu \overline{\overline{ \chi}}_\text{me} \sqrt{\frac{\epsilon}{\mu}}  \mathbf{E}_\text{av} + i \omega \mu \overline{\overline{ \chi}}_\text{mm} \mathbf{H}_\text{av}  ,
	\end{equation}
\end{subequations}
where the symbol $\Delta$ and the subscript `av' represent the differences and averages of the tangential fields at both sides of the metasurface, i.e.,

\begin{equation}\label{eq:field differences}
	\Delta\boldsymbol{\Phi} = \boldsymbol{\Phi}_b - \boldsymbol{\Phi}_a,
\end{equation}
\begin{equation}\label{eq:field averages}
	\boldsymbol{\Phi}_\text{av} = (\boldsymbol{\Phi}_a+\boldsymbol{\Phi}_b)/2,
\end{equation}

where $\boldsymbol{\Phi}=\mathbf{E},\mathbf{H}$, and $\overline{\overline{\chi}}_\text{ee}$, $\overline{\overline{\chi}}_\text{em}$, $\overline{\overline{\chi}}_\text{me}$ and $\overline{\overline{\chi}}_\text{mm}$ are the bianisotropic surface susceptibility tensors describing the metasurface. In this paper, we shall restrict our attention to purely tangential susceptibility metasurfaces, corresponding to $2\times{2}$ tensors and hence 16 susceptibility parameters in Equation~\eqref{eq:GSTC_tangential}, although metasurfaces involving normal susceptibility components may offer further possibilities~\cite{achouri2020electromagnetic}, as will be discussed later.

\idt We heuristically start our quest for the design described in connection with \figref{fig:Brewster_problem} by considering the simplest type of metasurface, namely an homoanisotropic metasurface, which is defined as a  metasurface whose only nonzero susceptibility tensors are $\overline{\overline{\chi}}_\text{ee}$ and $\overline{\overline{\chi}}_\text{mm}$. The nongyrotropy condition implies then $\chi_\text{ee}^{xy}=\chi_\text{ee}^{yx}=\chi_\text{mm}^{xy}=\chi_\text{mm}^{yx}=0$~\cite{achouri2020electromagnetic}, which decouples the two polarizations with $\chi_\text{ee}^{xx}$ and $\chi_\text{mm}^{yy}$ for TM and $\chi_\text{ee}^{yy}$ and $\chi_\text{mm}^{xx}$ for TE (see \figref{fig:Brewster_problem}(b)). Inserting the specifications~\eqref{eq:fields_TM} and~\eqref{eq:fields_TE} into the field differences and averages~\eqref{eq:field differences} and~\eqref{eq:field averages}, substituting the resulting expressions into into~\eqref{eq:GSTC_tangential}, and solving for the four nonzero susceptibility components yields
\begin{subequations}\label{eq:p-pol_mono_susceptibilities}
  \begin{equation}
\chi_\text{ee}^{xx} =\frac{2 i \eta_\text{a} T e^{i \phi_\text{TM} }-2 i \eta_\text{b}}{\eta_\text{a} \eta_\text{b} \omega  \epsilon_0 \cos \theta_\text{a}+\eta_\text{a} \eta_\text{b} T \omega  \epsilon_0 e^{i \phi_\text{TM} } \cos \theta_\text{b}},
\end{equation}
\begin{equation}
  \chi_\text{mm}^{yy} =-\frac{2 i \eta_\text{a} \eta_\text{b} \left(\cos \theta_\text{a}-T e^{i \phi_\text{TM} } \cos \theta_\text{b}\right)}{\mu_0 \omega  \left(\eta_\text{b}+\eta_\text{a} T e^{i \phi_\text{TM} }\right)}.
\end{equation}
\end{subequations}

\begin{subequations}\label{eq:s-pol_mono_susceptibilities}
\begin{equation}
\chi_\text{ee}^{yy} = \frac{2 i \eta_\text{a} T e^{i \phi_\text{TE} } \cos \theta_\text{b}-2 i \eta_\text{b} \cos \theta_\text{a}}{\eta_\text{a} \eta_\text{b} \omega  \epsilon_0+\eta_\text{a} \eta_\text{b} T \omega  \epsilon_0 e^{i \phi_\text{TE} }},
\end{equation}
  \begin{equation}
\chi_\text{mm}^{xx} = \frac{i \eta_\text{a} \eta_\text{b} \left(-1+T e^{i \phi_\text{TE} }\right)}{-\eta_\text{b} \mu_0 \omega  \cos \theta_\text{a}+\eta_\text{a} \mu_0 T \omega  e^{i \phi_\text{TE} } \cos \theta_\text{b}}.
\end{equation}
\end{subequations}

The complex nature of these susceptibilities indicates the presence of loss or gain, whereas we are searching for a lossless and gainless metasurface. This attempt is therefore unsuccessful, but it demonstrates the necessity for a really bianisotropic metasurface, as will be shown next.

\idt At this point, we can still hope that adding heteroanisotropy, corresponding to the susceptibilitity tensors $\overline{\overline{\chi}}_\text{em}$ and $\overline{\overline{\chi}}_\text{me}$, may allow to remove the loss-gain constraint of the previous design via the resulting extra degrees of freedom. Let us thus add the two heterotropic allowed pairs of nongyrotropic components, namely $\chi_\text{em}^{xy}$ and $\chi_\text{me}^{yx}$ for TM and $\chi_\text{em}^{yx}$ and $\chi_\text{me}^{xy}$ for TE. This increases the number of parameters to eight, and implies therefore the specification of an additional wave transformation for each polarization in order to make the system of equations~\eqref{eq:GSTC_tangential} full-rank and hence the synthesis problem exactly determined. Since some forms of bianisotropy can lead to nonreciprocity~\cite{caloz2018electromagnetic}, which is here prohibited, we shall enforce reciprocity by specifying a second wave transformation corresponding to the time-reversed version of the fields in~\eqref{eq:fields_TM} and~\eqref{eq:fields_TE}~\cite{jackson1999classical}. The resulting system of equations can be compactly written as \\
\begin{equation}\label{eq:TM_transf_syst}\hspace{-5mm}\vspace{-2mm}
\begin{bmatrix}
    \Delta H_{y1}    & \hspace{-3mm}     \Delta H_{y2} \\
    \Delta E_{x1}     & \hspace{-3mm}     \Delta E_{x2} \\
\end{bmatrix}
=
\begin{bmatrix}
    -i\omega\epsilon_0\chi^{xx}_\text{ee} &\hspace{-3mm} -ik_0\chi^{xy}_\text{em}  \\
    -ik_0\chi^{yx}_\text{me} &\hspace{-3mm} -i\omega\mu_0\chi^{yy}_\text{mm} \\
\end{bmatrix}
\begin{bmatrix}
          E_{x1,\text{av}} &\hspace{-3mm} E_{x2,\text{av}} \\
      H_{y1,\text{av}} &\hspace{-3mm} H_{y2,\text{av}}
\end{bmatrix},
\end{equation}

for the TM polarization, and as

\begin{equation}\label{eq:TE_transf_syst}
  \hspace{-5mm}
\begin{bmatrix}
    \Delta H_{x1}    & \hspace{-3mm}     \Delta H_{x2} \\
    \Delta E_{y1}     & \hspace{-3mm}     \Delta E_{y2} \\
\end{bmatrix}
=
\begin{bmatrix}
    -i\omega\epsilon_0\chi^{yy}_\text{ee}      &\hspace{-3mm} -ik_0\chi^{yx}_\text{em}  \\
    -ik_0\chi^{xy}_\text{me}       &\hspace{-3mm} -i\omega\mu_0\chi^{xx}_\text{mm} \\
\end{bmatrix}
\begin{bmatrix}
          E_{y1,\text{av}}      &\hspace{-3mm}   E_{y2,\text{av}} \\
      H_{x1,\text{av}} &\hspace{-3mm} H_{x2,\text{av}}
\end{bmatrix},
\end{equation}

\noindent for the TE polarization, where the subscript $1$ corresponds to the fields in~\eqref{eq:fields_TM} and~\eqref{eq:fields_TE}, and the subscript $2$ corresponds to their time-reversed counterpart. Solving this system for the eight susceptibility components yields
\begin{subequations}\label{eq:TM-pol_suscetibilities}
  \begin{equation}
\chi_\text{ee}^{xx} = -\frac{8 T \sin \phi_\text{TM} }{\omega  \epsilon  \left(T \alpha (\eta_\text{b} \cos \phi_\text{TM} +\eta_\text{a} T)+2 \cos \theta_\text{a} (\eta_\text{b}+\eta_\text{a} T \cos \phi_\text{TM} )\right)},
  \end{equation}
    \begin{equation}
\chi_\text{em}^{xy} = -\chi_\text{me}^{yx} = \frac{-2i  \left(T \alpha (\eta_\text{b} \cos \phi_\text{TM} -\eta_\text{a} T)+2 \cos \theta_\text{a} (\eta_\text{b}-\eta_\text{a} T \cos \phi_\text{TM} )\right)}{k \left(T \alpha (\eta_\text{b} \cos \phi_\text{TM} +\eta_\text{a} T)+2 \cos \theta_\text{a} (\eta_\text{b}+\eta_\text{a} T \cos \phi_\text{TM} )\right)},
  \end{equation}
    \begin{equation}
\chi_\text{mm}^{yy} = -\frac{8 \eta_\text{a} \eta_\text{b} T \cos \theta_\text{a} \sin \phi_\text{TM} \cos \theta_\text{b}}{u \omega  \left(T \alpha (\eta_\text{b} \cos \phi_\text{TM} +\eta_\text{a} T)+2 \cos \theta_\text{a} (\eta_\text{b}+\eta_\text{a} T \cos \phi_\text{TM} )\right)}
  \end{equation}
\end{subequations}
for the TM polarization and
\begin{subequations}\label{eq:TE-pol_suscetibilities}
  \begin{equation}
\chi_\text{ee}^{yy} = \frac{8 T \cos \theta_\text{a} \sin \phi_\text{TE}  \cos \theta_\text{b}}{\omega  \epsilon  \left(\eta_\text{a} T (T+\cos \phi_\text{TE} ) \alpha +2 \eta_\text{b} \cos \theta_\text{a} (T \cos \phi_\text{TE}+1)\right)},
  \end{equation}
    \begin{equation}
\chi_\text{em}^{yx} = -\chi_\text{me}^{xy} = -\frac{2 i \left(\eta_\text{a} T (T-\cos \phi_\text{TE} ) \alpha +2 \eta_\text{b} \cos \theta_\text{a} (T \cos \phi_\text{TE} -1)\right)}{k \left(\eta_\text{a} T (T+\cos \phi_\text{TE} ) \alpha+2 \eta_\text{b} \cos \theta_\text{a} (T \cos \phi_\text{TE} +1)\right)},
  \end{equation}
    \begin{equation}
\chi_\text{mm}^{xx} =  \frac{8 \eta_\text{a} \eta_\text{b} T \sin \phi_\text{TE} }{u \omega  \eta_\text{a} T (T+\cos \phi_\text{TE} ) \alpha+2 \eta_\text{b} \cos \theta_\text{a} (T \cos \phi_\text{TE} +1)}
  \end{equation}
\end{subequations}
for the TE polarization, with $\alpha =\sqrt{\frac{2 n_\text{a}^2 \cos (2 \text{$\theta $a})}{n_\text{b}^2}-\frac{2 n_\text{a}^2}{n_\text{b}^2}+4}$. These homotropic and heterotropypic susceptibilities are respectively purely real and purely imaginary, which indicates that the corresponding metasurface is lossless and gainless~\cite{achouri2020electromagnetic}. Thus, this design satisfies all the chosen requirements: it provides Brewster ($R=0$) transmission for arbitrary incidence and polarization while being lossless and gainless, nongyrotropic and reciprocal.

\begin{figure}[h]
	\centering
  \includegraphics[width=0.8\linewidth]{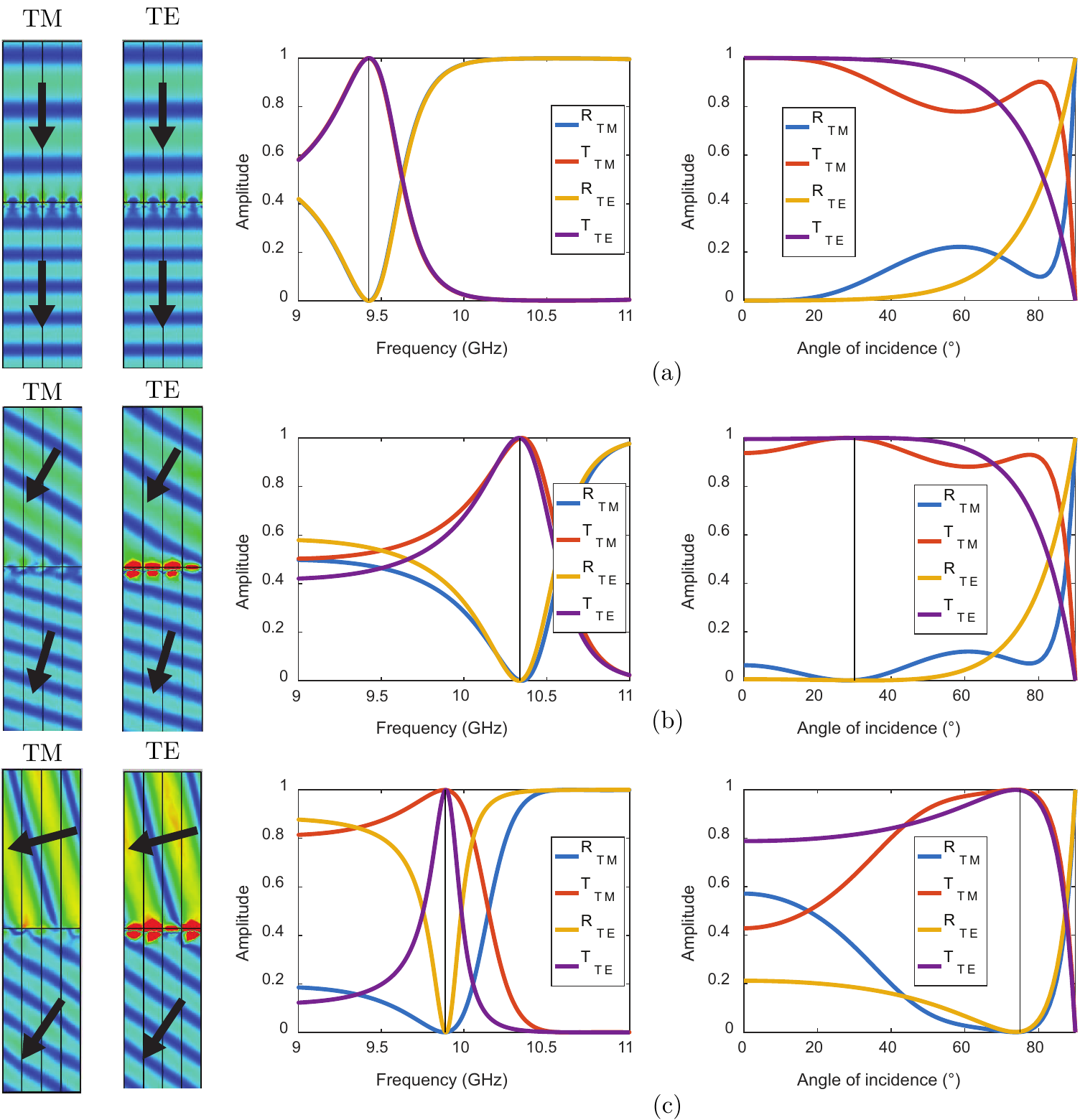}
	\caption{Full-wave simulated electric field amplitude distribution, frequency response and angular response of the reflectance and transmittance for polarization-independent $xz$-plane ($\varphi=0$) Brewster metasurfaces with the general parameters $(\epsilon_\text{r,a},\epsilon_\text{r,b})=(1,3)$ (bare-interface Brewster angle for the TM polarization at $60^\circ$),  $\epsilon_\text{r,subs}=3$, $d=1.52$~mm, $s = 0.5$~mm and $g=0.5$~mm. (a)~Brewster angle at $\theta_\text{a}=0$ (normal incidence) with $w_{y\text{u}}=w_{x\text{u}}=3.9$~mm and $w_{y\text{l}}=w_{x\text{l}}=2.75$~mm. (b)~Brewster angle at $\theta_\text{a}=30^\circ$ with $w_{y\text{u}}= 3.2$~mm, $w_{x\text{u}} =3.3$~mm, $w_{y\text{l}}= 2.2$~mm and $w_{x\text{l}} = 2.2$~mm. (c)~Brewster angle at $\theta_\text{a}=75^\circ$ with $w_{y\text{u}}= 3.35$~mm, $w_{x\text{u}} = 3.9$~mm, $w_{y\text{l}}= 2.65$~mm and $w_{x\text{l}} =2.45$~mm. }\label{fig:results2}
\end{figure}

\idt \figref{fig:results2} presents the results for the metasurface design with the susceptibilities~\eqref{eq:TM-pol_suscetibilities} and~\eqref{eq:TE-pol_suscetibilities}, which correspond to polarization-independent (TE and TM) Brewster transmission in the $xz$-plane. These results show that the specifications are perfectly realized by the designed metasurfaces for all the Brewster angles in the X-band frequency range and show also the angular response of the metasurface system around the Brewster angle design.

\idt The design of \figref{fig:results2}, with coinciding TM and TE Brewster angles, provides full reflection suppression for unpolarized light. However, this response is restricted to scattering in the $xz$ ($\phi=0$) plane. Indeed, according to Equations~\eqref{eq:TM-pol_suscetibilities} and~\eqref{eq:TE-pol_suscetibilities}, we have $\chi_\text{ee}^{yy}\neq\chi_\text{ee}^{xx}$, $\chi_\text{mm}^{xx}\neq\chi_\text{mm}^{yy}$, $\chi_\text{em}^{xy} \neq \chi_\text{em}^{yx}$ and $\chi_\text{me}^{xy}\neq\chi_\text{me}^{yx}$, and therefore the metasurface structure is anisotropic since the rotation $(x,y)\rightarrow(y,-x)$ implies different susceptibilities and different susceptibility cannot lead to the same scattering response.

\begin{figure}[h]
\centering
  \includegraphics[width=0.9\linewidth]{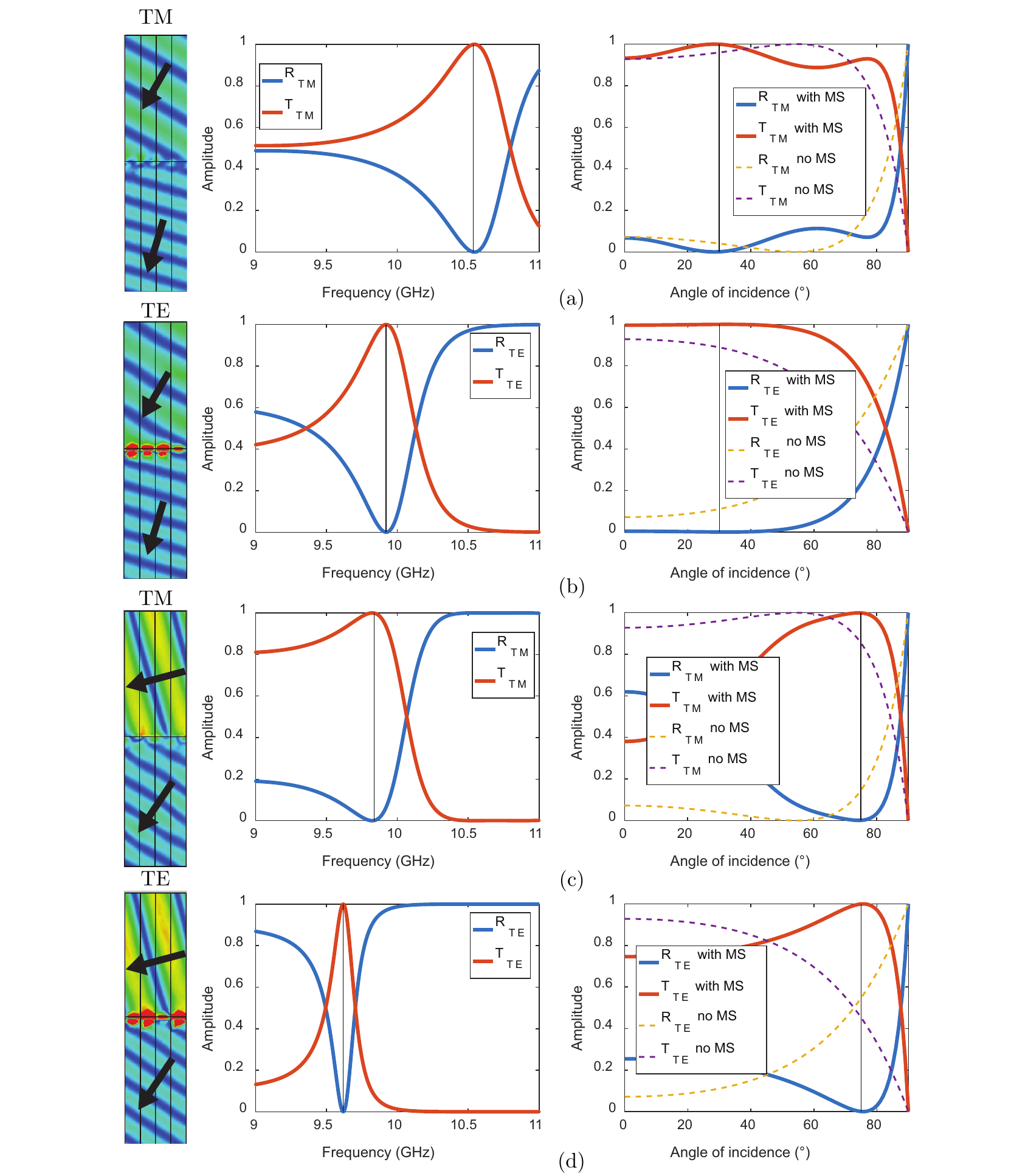}
\caption{Full-wave simulated electric field amplitude distribution, frequency response and angular response of the reflectance and transmittance for azimuth-independent ($\forall\varphi$) single-polarization (TM or TE) Brewster metasurfaces with the same general parameters as in \figref{fig:results2}. (a)~TM-Brewster angle at $\theta_\text{a}=30^\circ$ with $w_{x\text{u}}= w_{y\text{u}} = 3.15$~mm and $w_{x\text{l}}= w_{y\text{l}} = 2.1$~mm. (b)~TE-Brewster angle at $\theta_\text{a}=30^\circ$ with $w_{y\text{u}}= w_{x\text{u}} =3.3$~mm and $w_{y\text{l}}= w_{x\text{l}} = 2.45$~mm. (c)~TM-Brewster angle at $\theta_\text{a}=75^\circ$ with $w_{x\text{u}}= w_{y\text{u}} = 3.9$~mm and $w_{x\text{l}}= w_{y\text{l}} = 2.5$~mm. (d)~TE-Brewster angle at $\theta_\text{a}=75^\circ$ with $w_{y\text{u}}= w_{x\text{u}} = 3.55$~mm and $w_{y\text{l}}= w_{x\text{l}} = 2.85$~mm.}\label{fig:results1}
\end{figure}

\idt This single scattering plane restriction can be lifted with the same set of (eight) susceptibility parameters for one of the two polarizations (TM or TE) by combining the selected (TM or TE) $xz$-plane equations in \eqref{eq:TM-pol_suscetibilities} and~\eqref{eq:TE-pol_suscetibilities} with the corresponding $yz$-plane equations obtained via the permutations $(x,y)\rightarrow(y,-x)$, which is in fact equivalent to making the structure isotropic ($\chi_\text{ee}^{yy}=\chi_\text{ee}^{xx}$, $\chi_\text{mm}^{xx}=\chi_\text{mm}^{yy}$, $\chi_\text{em}^{xy}=\chi_\text{em}^{yx}$ and $\chi_\text{me}^{xy}=\chi_\text{me}^{yx}$) since the same Brewster response is expected in the two planes for the selected polarization. The results for corresponding metasurfaces are presented in~\figref{fig:results1}. They further confirm the accuracy of the proposed design.

\idt Although the eight-parameter metasurfaces considered here are restricted to either single-plane or single-polarization Brewster transmission, bianisotropic metasurfaces involving a greater number of susceptibility parameters might be able to provide universal Brewster transmission. Since the possibilities of transverse ($x$ and $y$) susceptibilities have been exhausted, such metasurfaces would require resorting to normal $z$ susceptibilities. Although the related design is in principle still analytically tractable thanks to the uniformity of the metasurface~\cite{achouri2020electromagnetic}, it is considerably more involved and will therefore be deferred to a later study.

\idt Equations~\eqref{eq:TM-pol_suscetibilities} and~\eqref{eq:TE-pol_suscetibilities} do not only provide the sought after Brewster transmission design. They point to an extra fundamental capability of an interfacing bianisotropic metasuface that occurs when $\phi_\text{TM}=\phi_\text{TE}=0$, which can be achieved by phase compensation or adjustement. In this case, we have $\chi_\text{ee}^{xx}=\chi_\text{mm}^{yy}=\chi_\text{ee}^{yy}=\chi_\text{mm}^{xx}=0$, which leads to the heteroanisotropic GSTCs
\begin{subequations}\label{eq:heteroanisotropic_GSTC}
	\begin{equation}
		\hat{z} \times \Delta\mathbf{H} =   j k \overline{\overline{ \chi}}_\text{em}   \mathbf{H}_\text{av},
	\end{equation}
	\begin{equation}
		\Delta \mathbf{E} \times \hat{z}   =   j k \overline{\overline{ \chi}}_\text{me}   \mathbf{E}_\text{av},
	\end{equation}
\end{subequations}
with $\overline{\overline{ \chi}}_\text{em}= -\overline{\overline{ \chi}}_\text{em}^\text{T}$ for reciprocity~\cite{caloz2018electromagnetic}. The corresponding reflection coefficients can easily be computed from general field expressions~\cite{achouri2020electromagnetic}. They read
\begin{subequations}
\begin{equation}\label{eq:modif_Fresnel}
	r_\text{TM} = \frac{\eta_1 \cos \theta_1 - \eta_\text{2,TM,eff} \cos \theta_2 }{\eta_1 \cos \theta_1 + \eta_\text{2,TM,eff} \cos \theta_2}
	\quad\text{and}\quad
	 \quad r_\text{TE} =\frac{\eta_\text{2,TE,eff} \cos \theta_1 - \eta_1 \cos \theta_2}{\eta_\text{2,TE,eff} \cos \theta_1  + \eta_1 \cos \theta_2},
\end{equation}
where
\begin{equation}\label{eq:eta_eff}
	\eta_\text{2,TM,eff} =\eta_2 \frac{(2i - k \chi_\text{em}^{xy})^2}{(2i + k \chi_\text{em}^{xy})^2}
	\quad\text{and}\quad
	\eta_\text{2,TE,eff} =\eta_2 \frac{(2i - k \chi_\text{em}^{yx})^2}{(2i + k \chi_\text{em}^{yx})^2}.
\end{equation}
\end{subequations}

\idt The relations~\eqref{eq:modif_Fresnel} have the same mathematical form as the conventional Fresnel reflection coefficients~\cite{saleh1991fundamentals}. This reveals that the proposed [medium -- bianisotropic metasurface -- medium] system is equivalent to a [medium -- \emph{effective medium}] system, with the effective medium having the impedances given by Equations~\eqref{eq:eta_eff}. Thus, inserting such a bianisotropic metasurface at the interface between two media or coating a dielectric medium exposed to free space with it can change the effective bulk impedance of the transmission medium, which enriches the design possibilities of existing materials.

\idt Although the proof of concept systems in Figures~\ref{fig:results2} and~\ref{fig:results1} pertain to the microwave regime, where bianisotropic metasurfaces (surrounded by air) have been well documented, bianisotropic metasurfaces have also been recently demonstrated in all-dielectric configuration~\cite{alaee2015all,odit2016experimental}. Therefore, the proposed concepts of metasurface-based generalized Brewster effective refractive medium can be readily extended to the optical regime, where they may be particularly beneficial in terms of reducing the insertion loss associated to impedance mismatch in many common components.

\idt In summary, we have shown that a properly designed bianisotropic metasurface placed at the interface between two dielectric media or coating a dielectric medium exposed to the air provides Brewster transmission at arbitrary angles and for both the TM and TE polarizations. We have presented a rigorous derivation of the corresponding surface susceptibility tensors and demonstrated the system by microwave proof-of-concept designs. Moreover, we have noted that such a system leads to the concept of effective refractive media with tailorable impedances. The proposed bianisotropic metasurfaces offer deeply subwavelength matching solutions for initially mismatched media, and alternatively lead to the possibility of on-demand manipulation of the conventional Fresnel coefficients. They represent thus a fundamental advance in optical science and posses a considerable potential for novel technological developments.

% References
\medskip

% Use the following code if you wish to generate your bibliography with BibTeX;
% replace the string "MSP-template" below with the name(s) of
% the BibTeX data base(s) you want to use.
% The resulting bibliography-output (the content of the .bbl file)
% must be pasted back into this file before submission.
% Please also include your BibTeX data base file(s) in your submission
% so that we can re-run BibTeX if necessary.
%
%\bibliographystyle{MSP}
%\bibliography{MSP-template}
\bibliographystyle{IEEEtran}
\bibliography{LIB}

\end{document}